\documentclass[10pt,conference]{IEEEtran}
\usepackage{cite}
\usepackage{amsmath,amssymb,amsfonts}
\usepackage{algorithmic}
\usepackage{graphicx}
\usepackage{textcomp}
\usepackage{xcolor}
\usepackage{dsfont}
\usepackage{hyperref}
\usepackage{url}
\usepackage{booktabs}
\usepackage{balance}
\usepackage[braket, qm]{qcircuit}
\usepackage{xspace}

\newcommand{\sys}{\textsc{Agent-Q}\xspace}

\begin{document}

%\title{\sys: Fine-Tuning Large Language Models on Quantum Optimization for Circuit Generation}

\title{\sys: Fine-Tuning Large Language Models for Quantum Circuit Generation and Optimization}

\author{\IEEEauthorblockN{Linus Jern}
\IEEEauthorblockA{\textit{Aalto University} \\
linus.jern@aalto.fi}
\and
\IEEEauthorblockN{Valter Uotila}
\IEEEauthorblockA{\textit{Aalto University} \& \textit{University of Helsinki}\\
%\textit{University of Helsinki}\\
valter.uotila@aalto.fi}
\and
\IEEEauthorblockN{Cong Yu}
\IEEEauthorblockA{\textit{Aalto University} \\
cong.yu@aalto.fi}
\and
\IEEEauthorblockN{Bo Zhao}
\IEEEauthorblockA{\textit{Aalto University} \\
bo.zhao@aalto.fi}
}

\maketitle
% for Arxiv
% \begingroup
% \renewcommand\thefootnote{}\footnotetext{
% Contact emails follow the format \texttt{\{first.last\}@aalto.fi}.
% }
% \addtocounter{footnote}{-1}
% \endgroup

\begin{abstract}
Large language models (LLMs) have achieved remarkable outcomes in complex problems, including math, coding, and analyzing large amounts of scientific reports. Yet, few works have explored the potential of LLMs in quantum computing. The most challenging problem is to leverage LLMs to automatically generate quantum circuits at a large scale. Fundamentally, the existing pre-trained LLMs lack the knowledge of quantum circuits. In this paper, we address this challenge by fine-tuning LLMs and injecting the domain-specific knowledge of quantum computing. \looseness=-1

We describe \sys, an LLM fine-tuning system to generate and optimize quantum circuits. In particular, \sys implements the mechanisms to generate training data sets and constructs an end-to-end pipeline to fine-tune pre-trained LLMs to generate parameterized quantum circuits for various optimization problems. \sys provides 14,000 quantum circuits covering a large spectrum of the quantum optimization landscape: 12 optimization problem instances and their optimized QAOA, VQE, and adaptive VQE circuits. Based thereon, \sys fine-tunes LLMs and constructs syntactically correct parametrized quantum circuits in OpenQASM 3.0. We have evaluated the quality of the LLM-generated circuits and parameters by comparing them to the optimized expectation values and distributions. Experimental results show superior performance of \sys, compared to several state-of-the-art LLMs and better parameters than random. \sys can be integrated into an agentic workflow, and the generated parametrized circuits with initial parameters can be used as a starting point for further optimization, \emph{e.g.,} as templates in quantum machine learning and as benchmarks for compilers and hardware.
\end{abstract}


\section{Introduction}

Large language models (LLMs) have shown increasing capabilities in various tasks. While originally designed for text generation, LLMs have since excelled in code generation \cite{deepseek-aiDeepSeekCoderV2BreakingBarrier2024} \cite{huangOpenCoderOpenCookbook2024} \cite{roziereCodeLlamaOpen} \cite{teamCodeGemmaOpenCode2024}, music generation \cite{agostinelliMusicLMGeneratingMusic2023}, and even image and video generation \cite{kohGeneratingImagesMultimodal2023} \cite{yangCogVideoXTexttoVideoDiffusion2025}. LLMs will also likely help support quantum algorithm developers and quantum computing end-users in various tasks such as quantum circuit generation, hybrid quantum-classical code generation, and circuit compilation.

Quantum computing requires deep expertise in quantum algorithms and hardware. For instance, in quantum machine learning and optimization problems, it is nontrivial to configure an optimal parametrized circuit that can be used to solve the given problem. Finding a performant initial starting point for the quantum optimization and training routines might be even harder. The convergence of optimization and training depends on the initial selection of circuit parameters. As a result, quantum circuit generation and parameter initialization appear to be promising problems for LLMs.

In this paper, we study how well LLMs can generalize to optimization problems in quantum computing.
To this end, we have designed \sys to fine-tune a pre-trained LLM with specially-crafted data sets that contain optimized QAOA, VQE, and adaptive VQE circuits for common optimization problems. Such optimization problems have been solved, producing circuits with optimized parameters -- making it one of the largest quantum circuit data sets of over 14,000 circuits. 

\sys also provides prompts to indicate the optimization problem and the corresponding optimized circuits. After fine-tuning, a user can design a prompt specifying an optimization problem and ask \sys to produce various circuits with initial parameters to solve the optimization problem. We show that the circuits generated by \sys's fine-tuned LLM outperform the state-of-the-art LLMs, and the initial parameters are often \emph{closer} to the optimal value than random ones. We evaluate the performance of \sys's model using three metrics: (i) syntactic correctness, (ii) ability to generate circuits with expectation values close to the optimized targets, and (iii) ability to produce circuits with probability distributions that align with the optimized ones.

\sys trains an LLM to produce pure quantum circuits instead of creating quantum-classical code. Focusing solely on quantum circuits, we introduced evaluation metrics that depend on solutions to the optimization problems, which makes the evaluation more robust. Current LLMs are very good at Python code but lack knowledge about quantum computing. Thus, it is likely that the models will learn the hybrid quantum-classical pipelines relatively easily after they understand quantum computing and classical computing separately.

Moreover, quantum circuit generation might be useful not only as an initial step for quantum optimization pipelines but also for other tasks. For example, circuit compilers, quantum error correction, and mitigation algorithms can be benchmarked with LLM-generated circuits. Additionally, the fact that LLMs could already generate framework-independent quantum circuits (\emph{e.g.}, Qiskit, PennyLane, Cirq) marks an important first step toward more advanced capabilities, where LLMs could help create larger hybrid workloads that seamlessly combine Python and quantum code.

We describe \sys, an end-to-end LLM fine-tuning framework to generate and optimize quantum circuits.
\sys makes the following new technical contributions:
\begin{itemize}
    \item A comprehensive dataset of over 14,000 quantum circuits, identifying 12 optimization with QAOA, VQE, and adaptive VQE and their optimized circuits, making it one of the largest collections of quantum circuits.
    \item An end-to-end LLM fine-tuning pipeline that automatically generates syntactically correct quantum circuits. 
    \item We have conducted comprehensive experiments to show \sys's fine-tuned LLM outperforms the state-of-the-art models and produces initial parameters that are closer to optimal than random.  We have open-sourced the code and data in \cite{github_repo, linuzj_graph_data_quantum}.
\end{itemize}

This article is organized as follows. First, we review the data set generation that consists of 12 optimization problems, which are used to create the training and test circuits with the optimized parameters. Then, we describe the fine-tuning pipeline, which produces a large language model fine-tuned on the previously generated data. Next, we present the evaluation, which covers syntax and performance metrics related to the optimization problems. In the discussion section, we analyze the results and suggest multiple possible next steps.

\subsection{Related work and previous use cases for generated circuits}

Only a few previous works have utilized large language models for quantum code generation. One of these works is \cite{dupuis2024qiskit}, which is the basis for Qiskit Code Assistant \cite{qiskit_code_assistant} and evaluated with \cite{vishwakarma2024qiskit}. These works are strongly built around Qiskit, while we designed our model to work with OpenQASM 3.0 \cite{cross_openqasm_2022}, which has become a platform-independent standard supported by Qiskit, Pennylane, and Cirq. Language models have also been used to design quantum experiments \cite{arlt2024meta}.

While LLMs have not yet been widely utilized in quantum computing, the standard transformer-based model has been used in various quantum computing applications. For example, the standard GPT model predicted measurement outcomes from a neutral atom quantum computer \cite{Fitzek_Teoh_Fung_Dagnew_Merali_Moss_MacLellan_Melko_2024}. The work showed how the standard GPT model has certain limitations when trained to predict measurement outcomes. These findings might be helpful to broaden our understanding of the limitations of the current LLM models.

Nvidia has developed a transformer-based optimization pipeline that generates quantum circuits in the search for ground states of electronic structure Hamiltonians \cite{Nakaji_Kristensen_Campos-Gonzalez-Angulo_Vakili_Huang_Bagherimehrab_Gorgulla_Wong_McCaskey_Kim_et_al_2024}. Since the trainable parameters are in the transformer model, the method aims to circumvent specific problems that the current variational methods have, such as barren plateaus.

KetGPT uses GPT-based models to generate quantum circuits \cite{Apak_Bandic_Sarkar_Feld_2024}, which have been trained on QASMBench circuits \cite{10.1145/3550488}. The produced synthetic circuits mimic the structure in the training dataset. The circuits are limited to OpenQASM 2.0 format, without supporting parameters. \looseness=-1
\section{Quantum optimization data set for fine-tuning}
\label{sec:data}

The training dataset is one of the most essential components in fine-tuning large language models. This section describes how we have constructed a large, high-quality, and diverse data set consisting of around 14,000 optimized quantum circuits for multiple key optimization primitives on graphs commonly used in classical and quantum optimization algorithms. As a result, we have not only constructed a comprehensive training data set but also a data set consisting of circuits that can be alternatively used for hardware and circuit compilation benchmarking and other tasks. The circuits are expressed in the most recent OpenQASM 3.0 format, which supports parametrized quantum circuits, and the dataset is available on HuggingFace \cite{linuzj_graph_data_quantum}, and the code to generate the dataset is available on GitHub \cite{github_repo}.

As pointed out in \cite{Muennighoff_Yang_Shi_Li_Fei_Fei_Hajishirzi_Zettlemoyer_Liang_Candes_Hashimoto_2025}, the key features we want from the dataset are quality, difficulty, and diversity. Considering the dataset we have created, we have aimed to satisfy these key characteristics:
\begin{itemize}
    \item Quality: The data not only contains circuits for optimization problems but also the optimized parameters. If the circuits are executed with the given parameters, there is a high probability of measuring the bitstring corresponding to the correct solution.
    \item Difficulty: Creating high-performing quantum circuits with good parameters is known to be a challenging task.
    \item Diversity: We have included 12 different optimization problems on graphs. Moreover, we have solved the problems with QAOA, VQE, and adaptive VQE.
\end{itemize}

\subsection{Optimization problems}

This subsection briefly reviews the implemented optimization problems and our optimization methods. Since the goal is to generate high-quality parametrized quantum circuits for optimization problems, the training dataset should contain a representative set of such problems. The optimization problems we have considered are standard primitives in various classical optimization algorithms, and many of them appeared in \cite{Lucas_2014}. We have limited our focus on optimization problems on graphs, which form the core of optimization algorithms \cite{Karp_1972}. The selected problems are listed in \autoref{tab:dataset_problems}, and their implementations are on GitHub~\cite{github_repo}.

\begin{table}[ht]
    \centering
    \caption{Optimization problems in the training dataset. The complexity refers to the classical decision variant. The binary optimization formulations for problems without reference are developed in this work.}
    \resizebox{\columnwidth}{!}{%
        \begin{tabular}{lcl}
        \toprule
        \textbf{Problem}
            & \textbf{Formulation}
            & \textbf{Classical complexity} \\
        \midrule
        Connected Components           & QUBO & P \cite{intro_to_alg_2009} \\
        Community Detection \cite{negre_2020_detecting}            & QUBO & NP-hard \cite{fortunato_2016}\\
        $k$-Clique \cite{Lucas_2014}                    & QUBO & NP-complete \cite{Karp_1972}\\
        Graph Isomorphism~\cite{Lucas_2014}              & QUBO & in NP (open) \cite{babai_gi_2016, Johnson_1987} \\
        Graph Coloring~\cite{Lucas_2014}                  & QUBO & NP-complete \cite{Karp_1972} \\
        Traveling Salesman~\cite{Lucas_2014}             & QUBO & NP-complete \cite{Karp_1972} \\
        Weighted Minimal Maximal Matching & QUBO & NP-Hard \cite{Lucas_2014} \\
        Vertex Cover~\cite{Lucas_2014}                   & QUBO & NP-complete \cite{Karp_1972} \\
        Edge Cover                      & HUBO & P \cite{computer_intractably_1990} \\
        Max-Flow~\cite{9224181}                      & QUBO & P \cite{max_flow_2013} \\
        Min-Cut~\cite{9224181}                        & QUBO & P \cite{min_cut_1994} \\
        (Hyper)MaxCut \cite{farhiQuantumApproximateOptimization2014}                    & HUBO & MaxCut NP-complete \cite{Karp_1972} \\
        \bottomrule
        \end{tabular}
    }
    \label{tab:dataset_problems}
\end{table}

%We have also constructed a quantum formulation for a new problem: MaxCut on hypergraphs. We will explain this problem in detail in the following section. Before that, we briefly introduce the other problems and their quantum formulations.

\subsubsection{Connected components in graphs}
Finding connected components in a graph $G$ means finding a partition $P$ of $G$ such that every subgraph in $P$ is connected, meaning that every two nodes in the subgraph are connected with a path. In this implementation, we consider a Quadratic Unconstrained Binary Optimization (QUBO) formulation for the connected components problem. In this formulation, we fix a node in a graph, and the algorithm returns the connected component to which the fixed node belongs.

\subsubsection{Community detection} 
In the community detection problem, the goal is to find a partition $P$ of a graph $G$ so that the density of the edges within the partitions in $P$ is higher than the density of edges between them. In this work, we implemented the QUBO formulation for the community detection algorithm based on the modularity measure, which describes the quality of the partition into communities \cite{clauset_2004, negre_2020_detecting}. The problem is proved to be NP-hard \cite{fortunato_2016}.

\subsubsection{k-sized clique}
The QUBO formulation for finding $k$-sized clique was developed in \cite{Lucas_2014}. The problem is to find the complete subgraph of size $k$ from a given graph. The decision problem of whether a $k$-sized clique exists is NP-complete \cite{Karp_1972}.

\subsubsection{Graph isomorphism}
Graph isomorphism is the problem of determining if there exists a bijective mapping $f \colon V_1 \to V_2$ between the vertex sets of graphs $G_1$ and $G_2$ such that whenever $(v_1, v_2) \in E_1$ is an edge in graph $G_1$, then $(f(v_1), f(v_2)) \in E_2$ is an edge in graph $E_2$. Interestingly, graph isomorphism is known to be in NP, but it is unknown if it is NP-complete \cite{Johnson_1987}. In practice, it is a complex problem. We have implemented the standard QUBO formulation for graph isomorphism~\cite{Lucas_2014}, but formulations also exist for adiabatic quantum computers \cite{Gaitan_2014} and boson samplers \cite{Bradler_2021}.

\subsubsection{Graph coloring}
Given $n$ colors and a graph $G$, the graph coloring problem is to determine if the $n$ colors can be assigned to the vertices of $G$ so that no edge connects two vertices of the same color. The problem is known to be NP-complete \cite{Karp_1972}. We implemented the QUBO formulation from \cite{Lucas_2014}.\looseness=-1

\subsubsection{Traveling salesman}
The traveling salesman problem is the optimization problem where, starting from a given node, the goal is to find a path in a weighted graph that visits every node in the graph exactly once. We implement the formulation from \cite{Lucas_2014}. The decision problem is NP-complete \cite{Karp_1972}.

\subsubsection{Weighted minimal maximal matching}
Minimal maximal matching is a special case of matching on graphs. A matching in a graph $G$ is a subset of its edges such that no two edges are adjacent to the same vertex. Matching problems generally are not NP-hard \cite{Edmonds_1965, Edmonds_1965_2} without additional constraints requiring minimality over the selected edges \cite{Lucas_2014}. A maximal matching is such a solution that if any edge that is not yet in the matching is included, the subset of edges would not be a matching anymore. In this work, we consider the problem of finding a maximal matching on a weighted graph with the minimum cost \cite{Cook_Rohe_1999}. The algorithm returns a perfect matching or a near-perfect matching when they exist, since these matchings are automatically maximal. This problem was formulated in \cite{Lucas_2014}, but we consider it as a special instance of the exact set cover, where we identify the edges on the graph with two-element sets. This way, we can use the exact set cover formula in \cite{Lucas_2014}, simplifying the formulation.

\subsubsection{Vertex cover}
The vertex cover problem seeks the smallest set of vertices in a graph such that every edge has at least one endpoint in this set. Our QUBO formulation is based on \cite{Lucas_2014}, and QAOA was previously benchmarked on this problem \cite{Cook_2019}. The decision version of the problem is NP-complete~\cite{Karp_1972}. 

\subsubsection{Edge cover}
The edge cover is similar to the vertex cover problem, but expressed for edges: what is the smallest set of edges such that every vertex in the graph is adjacent to at least one edge in this set? This problem is no longer NP-hard, as we can utilize the maximum matching algorithm to find a matching that can be greedily extended to form an edge cover. We did not find a standard QUBO formulation for this problem. Hence, we present a new higher-order formulation for it, which is inspired by the formulation for the vertex cover problem \cite{Lucas_2014}. We define $|E|$ many binary variables $x_e$ for each edge $e \in E$. If $x_e = 1$, the corresponding edge $e$ belongs to the covering. The first part of the Hamiltonian becomes
\begin{equation*}
    H_A = A\sum_{v \in V}\sum_{e \in N(v)}(1 - x_{e}).
\end{equation*}
The Hamiltonian $H_A$ encodes that we must select at least one edge for each vertex. Hamiltonian $H_A$ is a higher-order polynomial because the neighbor set $N(v)$ can generally contain more than $2$ elements. Then, we encode the cost with the standard
\begin{equation*}
    H_B = B\sum_{e \in E}x_{e}.
\end{equation*}
As in the case of the vertex cover problem, we require $A > B$. The second way to encode the edge cover problem is to use the inequality constraint methods in \cite{Lucas_2014}, but these methods require a logarithmic number of slack variables. In this formulation, we do not need the slack variables, which makes it more scalable in terms of required qubits.

\subsubsection{MaxFlow}
A flow network is a directed graph with source and sink nodes so that each edge has a non-negative capacity. The network does not have self-loops. A flow in this graph is a function $f \colon E \to \mathds{R}$ that assigns a real value $f(u, v)$ for each edge $(u, v) \in E$ representing the amount of flow. A maximum flow problem seeks to find a feasible flow from the source to the sink through the flow network, obtaining the maximum flow rate. The QUBO formulation for the MaxFlow problem was developed in \cite{9224181}.

\subsubsection{MinCut}
In the same work \cite{9224181}, where a QUBO formulation for the MaxFlow problem was introduced, the authors also developed a QUBO formulation for the MinCut problem. MinCut is complementary to MaxCut, as it involves minimizing the cut rather than maximizing it.

\subsubsection{MaxCut on hypergraphs}
While the other problems are well-known optimization problems, MaxCut on hypergraphs (HyperMaxCut) has not been previously proposed as a quantum optimization problem. We identified that it is well-suited for the data generation task because it has characteristics similar to MaxCut on graphs, which is one of the most studied optimization problems in quantum computing \cite{farhiQuantumApproximateOptimization2014, Basso_Farhi_Marwaha_Villalonga_Zhou_2022, Guerreschi_Matsuura_2019, Herrman_Treffert_Ostrowski_Lotshaw_Humble_Siopsis_2021, Majumdar_Madan_Bhoumik_Vinayagamurthy_Raghunathan_Sur-Kolay_2021, Rehfeldt_Koch_Shinano_2023, Wang_Hadfield_Jiang_Rieffel_2018, Zhou_Du_Tian_Tao_2023}. The key difference between MaxCut and HyperMaxCut is that HyperMaxCut creates higher-order binary optimization problems, which means that we have terms with more than just two interacting variables. While these problems naturally map to quantum circuits, state-of-the-art classical solvers, such as Gurobi and CPLEX, do not natively support them.

Next, we define MaxCut for hypergraphs \cite{Conlon_Fox_Kwan_Sudakov_2019}. Let $\mathcal{G} = (V, E)$ be an undirected hypergraph, i.e., simply a set of nodes $V$ and a set of edges $E \subset \mathcal{P}(V)$ where $\mathcal{P}(V)$ is the powerset of $V$. We obtain the standard graph if we restrict $|e| = 2$ for all $e \in E$. We define MaxCut on hypergraphs, called HyperMaxCut, analogously to MaxCut on graphs by seeking a partition $z$ (i.e., two sets $A$ and $B$) of the vertex set $V$ in such a way that
\begin{displaymath}
C(z) = \sum_{i = 1}^{|E|}C_{i}(z),
\end{displaymath}
is maximized. In this formulation, $C_{i}(z) = 1$ if the solution $z$ places at least one vertex in the edge $e_{i}$ to $A$ and the others to the set $B$ (meaning we cut the edge $e_{i}$). Otherwise, $C_{i}(z) = 0$, which also means that every vertex in edge $e_{i}$ belongs to either $A$ or $B$. In the case that $|e| = 2$ for all $e \in E$, this formulation reduces to the standard MaxCut on graphs \cite{farhiQuantumApproximateOptimization2014}.

While the definition is a MaxCut generalization, there are other ways to formulate MaxCut for hypergraphs \cite{Conlon_Fox_Kwan_Sudakov_2019}. We chose this definition because it involves the partition of two sets ($A$ and $B$), which is easy to encode in binary and spin systems.

Next, we describe the higher-order binary optimization problem formulation for HyperMaxCut. Let us assume that the vertices admit a natural order indexed by $i \in [|V| - 1]$. We create a set of spin variables $Z = \left\{z_i \in \left\{-1,1 \right\} \mid v \in V \right\}$, which have the interpretation that $z_i = -1$ if $v_i \in A$ and otherwise $v_i \in B$. Thus, there is a clear correspondence between naturally ordered spin-strings, corresponding variables, and partitions of the hypergraph. Moreover, this correspondence is bijective. We use spin variables since the MaxCut problem on graphs notoriously has a simple formulation in terms of these variables in contrast to binary variables \cite{farhiQuantumApproximateOptimization2014}.

Let $e_{ij}$ be an edge between nodes $i$ and $j$. Considering MaxCut on graphs, the idea is to express $C_{e_{ij}}(Z)$ in such a way that the Hamiltonian corresponding $C_{e_{ij}}(Z)$ achieves its minimum at eigenstates $|00\rangle$ and $|11\rangle$ because in these cases $z_i = z_j$ which means that there is no cut between vertices $i$ and $j$. Thus, maximizing such a Hamiltonian will obtain a partitioning $Z$, which maximizes cuts between the vertices.

For the standard MaxCut on graphs ($|e_i| = 2$ for all $i$), the corresponding Hamiltonian for each edge is:
\begin{displaymath}
C_{e_{ij}}(Z) = -z_i z_j.
\end{displaymath}
Every edge set contains two vertices producing exactly $|E|$ many quadratic terms: $z_iz_j$ for every $e_{ij} \in E$ between $v_i$ and $v_j$. Utilizing the standard rewriting technique between spin and binary variables, we obtain the equivalent QUBO:
\begin{displaymath}
C_{e_{ij}}(X) = -(1 - 2x_i)(1 - 2x_j) = x_i + x_j - 2x_ix_j.    
\end{displaymath}
The simple evaluation at points $00$, $01$, $10$ and $11$, gives us values $0$, $1$, $1$ and $0$, respectively. The previous reasoning generalizes to hypergraphs and defines HyperMaxCut.

For a hyperedge $e\in E$, we want to define a formulation for $C_{e}(Z)$. Again, the function should achieve its minimum at states $|00 \ldots 0 \rangle$ and $|11 \ldots 1\rangle$. Let $n := |e|$ and if $n$ is even, let $m = n$ and otherwise $m = n - 1$. Then, the cost Hamiltonian in terms of spin variables is
\begin{align*}
C_{e}(Z) =& 
\frac{2^{n - 2} - 1}{2^{n - 2}} I - \frac{1}{2^{n - 2}}\sum_{\left\{ i, j \right\} \subset e} z_iz_j + \ldots \\
 &- \frac{1}{2^{n - 2}}\sum_{M \subset e, |M| = m}\prod_{i \in M}z_i,
\end{align*}
where the sum contains every even-length spin variable combination up to $|e|$. Thus, we obtain a higher-order optimization problem, which always contains terms of even order. This formulation can be used to solve the HyperMaxCut problem.

\subsection{Problem generation for optimization}

The optimization algorithms previously described cannot be utilized unless we generate optimization problem instances. The problem instances are graphs, and in many cases, they also include additional problem-specific parameters. For example, the $k$-sized clique problem requires a value for $k$. Thus, for each graph algorithm, we have constructed a problem instance generator, which constructs graphs to which the algorithms can be applied. For example, for the community detection algorithm, the generator creates graphs with reasonable communities that can be detected. For the graph isomorphism problem, the generator generates a graph and an automorphism, which is then applied to the first graph to obtain a pair of isomorphic graphs. For $k$-sized cliques, the generator generates a graph that is ensured to contain a $k$-sized clique. This method is necessarily a reverse-engineered approach to problem generation. 

\subsection{Optimization methods}

To solve the optimization problems using the quantum formulations for the graph algorithms, we have employed three standard optimization methods from quantum computing: QAOA \cite{farhiQuantumApproximateOptimization2014}, VQE \cite{Peruzzo_McClean_Shadbolt_Yung_Zhou_Love_Aspuru-Guzik_OBrien_2014}, and adaptive VQE \cite{Grimsley_Economou_Barnes_Mayhall_2019}. In this subsection, we briefly describe the methods and the choices that we have made regarding each algorithm. The code is primarily written in Pennylane \cite{bergholm2022pennylaneautomaticdifferentiationhybrid}, but a portion of the code relying on OpenQASM 3.0 transformation is based on Qiskit, as QASM is a standard developed by IBM.

\subsubsection{QAOA} QAOA is one of the most common quantum optimization algorithms. A QAOA circuit consists of cost and mixer layers that are applied repeatedly. The cost layer is constructed based on the Hamiltonian, whose minimum eigenvalue we are solving. The mixer layer we used in this work is the standard $x$-mixer, i.e., the layer of parametrized $R_x$ rotations. The number of layers varied between 1 and 4.

\subsubsection{VQE} The VQE optimization routines require ansatzes \cite{Peruzzo_McClean_Shadbolt_Yung_Zhou_Love_Aspuru-Guzik_OBrien_2014}. We have implemented the well-researched set of ansatzes from \cite{Sim_Johnson_Aspuru_Guzik_2019} and especially focused on the ansatzes having identifiers $1$, $3$, $4$, $5$, $6$, $7$, $9$, $10$, $11$, $12$, $13$, $14$, $15$, $16$, $18$ which refer to the identifiers in \cite{Sim_Johnson_Aspuru_Guzik_2019}. The final full ansatz circuit results from multiple layers of these ansatzes. The number of layers varied between $1-4$. The different ansatz layouts were not mixed, although that might be a feasible method to extend the data set further.

\subsubsection{Adaptive VQE} 
\begin{figure*}[t]
    \centering
    \input{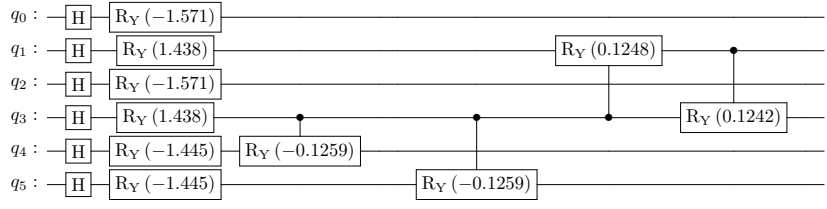}
    \caption{A short example circuit from the adaptive VQE algorithm}
    \label{fig:adaptive_vqe_example}
\end{figure*}

The adaptive VQE algorithm adds gates from a fixed pool adaptively depending on the circuit's gradient. Our implementation relies on the Pennylane implementation of the algorithm \cite{PennyLaneAdaptiveOptimizer}. The pool of gates for this implementation consists of a single qubit rotation $R_x$, $R_z$, and $R_y$, as well as the two-qubit gates that are the controlled versions of the single qubit rotations. Hence, the pool has six gate types, which can be positioned anywhere in the circuit with an optimized rotation. The application of the adaptive method started from the uniform superposition. An example of an adaptive circuit is visualized in \autoref{fig:adaptive_vqe_example}.

\textit{Stopping criteria.} The optimization was interrupted if the most probable solution from the quantum circuit was among the correct solutions from the exact eigensolver. The probabilities were computed analytically without errors. If the optimization did not converge, the case was classified as unsuccessful and not included in the training data.

\begin{figure}
    \centering
    \includegraphics[width=0.99\columnwidth]{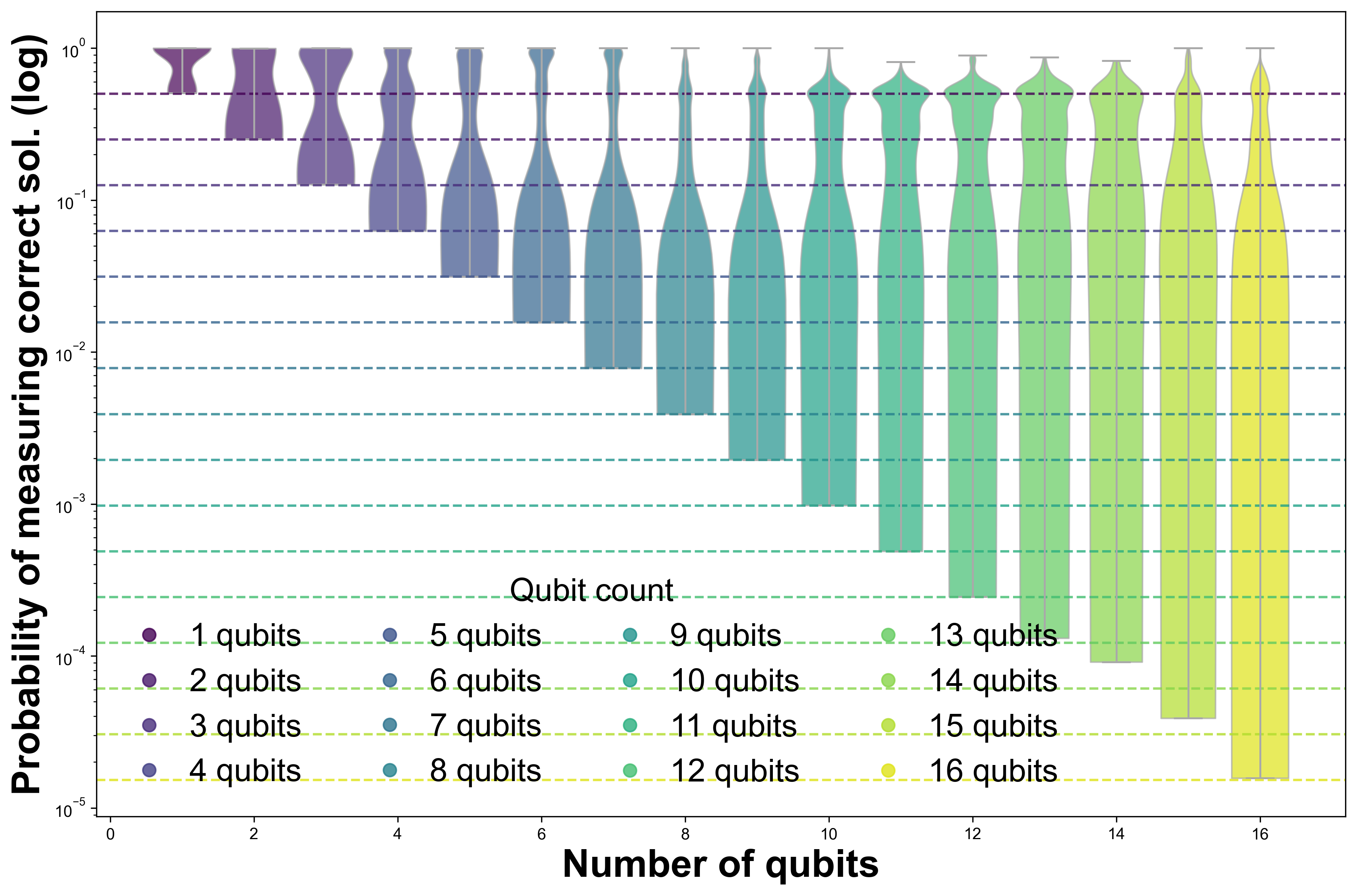}
    \caption{Distribution of probabilities to measure the correct solution after optimization grouped by the number of qubits. Each dashed line represents the uniform distribution that would be obtained without optimization. Since most circuits are top-of-the-line, we will likely measure the correct solution, indicating that the circuits are of high quality.}
    \label{fig:optimized_circuits_probability_distributions}
\end{figure}

\subsection{Data characteristics}

The attributes we have collected for supervised fine-tuning are as follows.
\begin{itemize}
    \item Hamiltonian encoding the optimization problem. Since there is no standard code-level notation for Hamiltonians, we decided to use Pennylane since this notation is also human-readable \cite{PennyLaneLinearCombination}.
    \item Smallest eigenvalue(s) and the first excited state(s) solved exactly with eigensolvers
    \item Total number of optimization steps to reach a sufficiently high probability to measure the correct solution
    \item The states with the highest probabilities
    \item QAOA, VQE, or adaptive VQE circuits with numeric and symbolic parameters
    \item For adaptive VQE, the circuits that have been created during the optimization process
    \item Problem-specific details (e.g., problem graph and additional parameters)
    \item Number of qubits, layers, and ansatz identifiers
\end{itemize}

The main output from the data generation process is the QAOA, VQE, and adaptive VQE circuits with optimized parameters. To make this system independent of the underlying Python framework, such as Qiskit, Pennylane, or Cirq, we decided to utilize circuits in the most recent OpenQASM 3.0 format \cite{Cross_Javadi_Abhari_Alexander_Beaudrap_Bishop_Heidel_Ryan_Sivarajah_Smolin_Gambetta_et_al_2022}. 

\begin{figure}[t]
    \centering
    \includegraphics[width=\columnwidth]{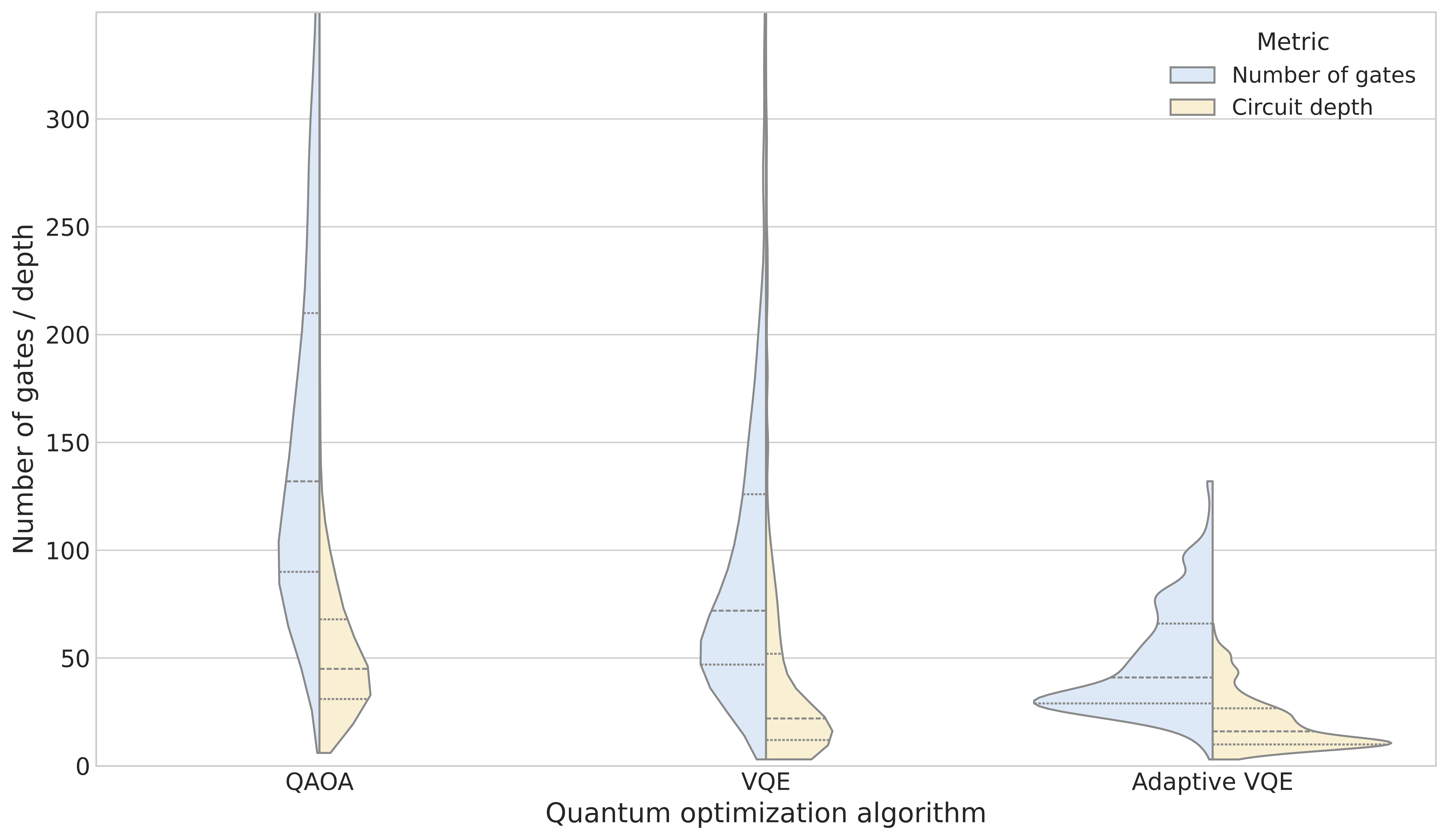}
    \caption{Distributions for number of gates and circuit depths for the training circuits grouped by the algorithm that has been used to solve the optimization problems}
    \label{fig:circuit_stats_violin}
\end{figure}

\begin{figure}[t]
    \centering
    \includegraphics[width=0.99\columnwidth]{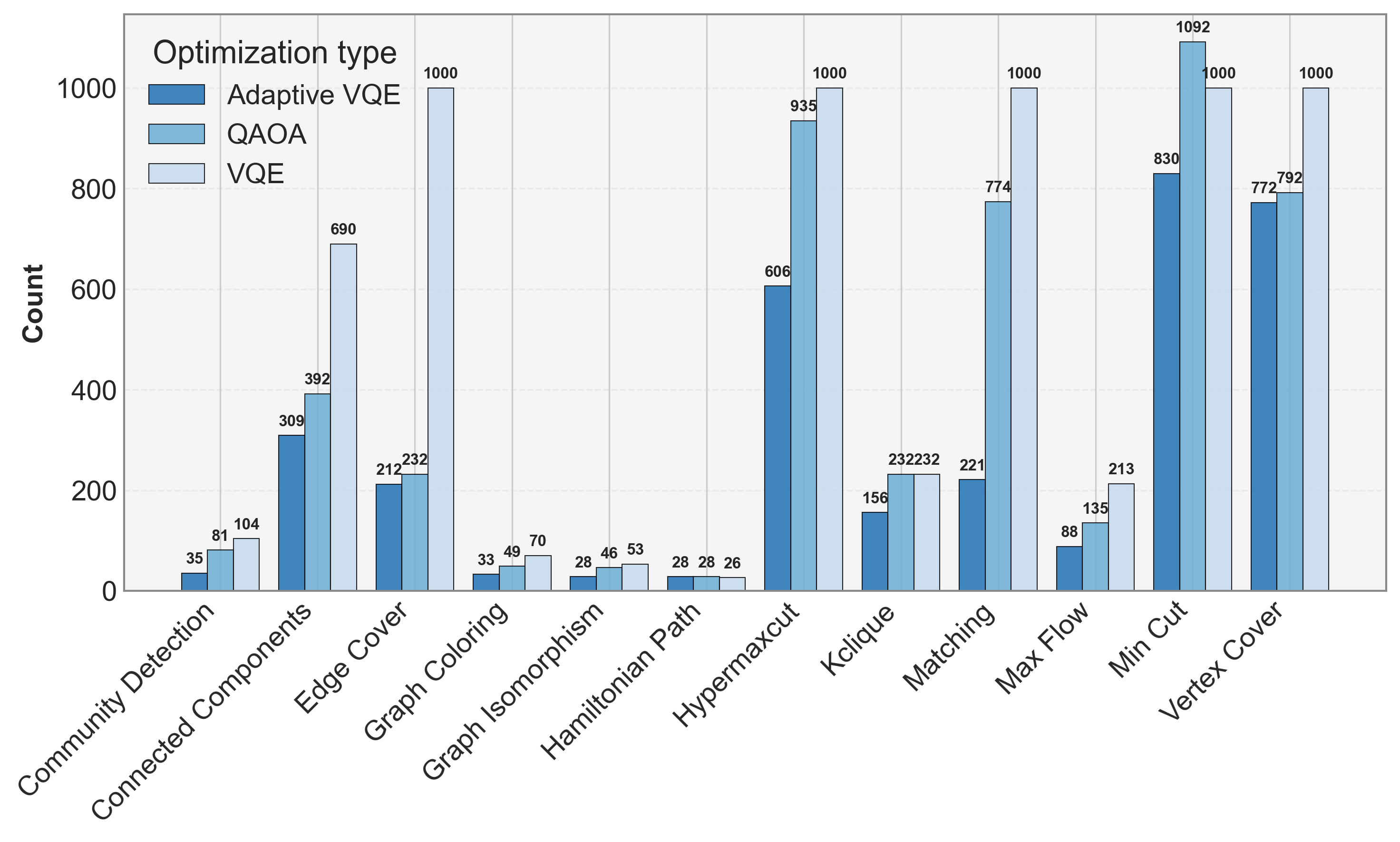}
    \caption{Number of problems and methods for each optimization problem}
    \label{fig:problem_distribution}
\end{figure}

\autoref{fig:circuit_stats_violin} maps the problems with respect to their qubit counts. We can see that the circuits are sufficiently optimized because, in the vast majority of cases, we have a relatively high probability (y-axis) of measuring the correct bit string. The thickness of the bars indicates the number of problems, and the dashed line is the probability of selecting a bitstring uniformly at random. As a summary, \autoref{fig:problem_distribution} displays the counts for different problems, grouped by the methods used.

%Currently, the implementation is limited to these problems because they can be described efficiently and admit relatively well-defined representation. For example, consider another prototypical optimization problem in quantum computing: portfolio optimization. In portfolio optimization, the problem is based on real-life stock market data, and the problem encodes interactions between every two variables. This leads to binary optimization problems, where the objective function is exponentially long with respect to the data, and the terms in the objective function have long floating-point coefficients. Inputting this type of data to the fine-tuning pipeline is computationally expensive. Thus, at the first stage, the problem set is limited to the optimization problems that are relatively efficient to encode and whose coefficients are integers or relatively short rational numbers.

\section{\sys's Fine-Tuning Pipeline}

% \todo[color=green!30,inline]{We should assume that the audience (quantum computer scientists) does not know anything about building and fine-tuning LLMs.

% Look at good examples: the deepseek paper, s1 paper

% Setup: pre-trained model, hyperparameters, resources (OpenAI papers good example?)

% Supervised fine-tuning

% Process diagrams: data generation and fine-tuning
% }

% Content:
% Introduce transformers/LLMs.

% Make clear distinction between pre-training and fine-tuning.

% Explain fine-tuning and different ways to do it.

% Explain our fine-tuning pipeline.

In recent years, transformer models, introduced in the groundbreaking paper \cite{vaswaniAttentionAllYou2023}, have reshaped the landscapes of natural language processing. The precision and flexibility of large language models based on transformer architecture have enabled breakthroughs in many fields, including code generation \cite{deepseek-aiDeepSeekCoderV2BreakingBarrier2024,huangOpenCoderOpenCookbook2024,roziereCodeLlamaOpen,teamCodeGemmaOpenCode2024}, music generation \cite{agostinelliMusicLMGeneratingMusic2023}, and even image and video generation \cite{kohGeneratingImagesMultimodal2023,yangCogVideoXTexttoVideoDiffusion2025}.
These models have been shown to be very capable of learning the relations within large amounts of sequential data, essentially learning patterns and logic from all publicly available data. 
In addition, scaling the model size has been shown to predictably improve the performance of these models via a phenomenon called emergent abilities \cite{weiEmergentAbilitiesLarge2022}.
This section will discuss the background of fine-tuning large language models and our approach to building a pipeline for generating quantum circuits.
We use the open-source state-of-the-art pre-trained model Qwen 2.5 Instruct \cite{qwenQwen25TechnicalReport2025}, trained by the Qwen team at Alibaba Cloud, as a base model that we fine-tune to generate quantum circuits.

\subsection{Transformer}

Transformers are a class of neural network architectures that have revolutionized sequence modeling and generation by generating tokens without relying on traditional recurrent structures.
Instead of relying on the traditional recurrence networks, they use a self-attention mechanism, allowing each element of the input sequence to interact with each other simultaneously, effectively capturing long-range dependencies \cite{vaswaniAttentionAllYou2023}.\looseness=-1

At the core of the transformer architecture, there are two components: the encoder and the decoder.
The encoder maps an input sequence of symbol representations $(x_1, \dots, x_n)$ to a sequence of continuous representaions $\boldsymbol{z}= (z_1, \dots, z_n)$.
Given the sequence $\boldsymbol{z}$, the decoder sequentially generates an output sequence $(y_1, \dots, y_m)$.
Throughout the generation process, each step is auto-regressive \cite{gravesGeneratingSequencesRecurrent2014}, meaning that all previously generated symbols are additional input when generating the next symbol.

% \begin{figure*}
%     \centering    \includegraphics[width=0.99\textwidth]{figures/transformer_with_residuals_nx_svg.png}
%     \caption{Transformer architecture as presented in \cite{vaswaniAttentionAllYou2023}}
%     \label{fig:transformer}
% \end{figure*}

To construct deep models, the encoder and decoder are organized into stacks of $N$ identical layers. 
These stacks form the \textit{Encoder stack} and the \textit{Decoder stack}.
Each layer within these stacks is connected via residual connections, which helps preserve the original signal and ensures stable gradients while enabling the increase in depth of the network, followed by layer normalization to improve convergence \cite{heDeepResidualLearning2015a} \cite{baLayerNormalization2016}.\looseness=-1

Each encoder block consists of two sub-layers.
The first layer implements a multi-head self-attention mechanism, and the second is a fully connected feed-forward neural network.
In simple terms, the output of each encoder block is given by: $\mathrm{LayerNorm}(x + \mathrm{Sublayer}(x))$, where $\mathrm{Sublayer}(x)$ is the function implemented within the encoder block.

The critical innovation of the transformer architecture is the multi-head attention mechanism. 
This mechanism effectively enhances the model's ability to capture different aspects of the input by dividing the input into multiple "heads".
For each head, the mechanism performs the following steps. 

\textit{Projection}: The input sequence is projected linearly to generate \textit{query}, \textit{key} and \textit{value} vectors. Let these vectors be $Q$, $K$, and $V$.

\textit{Attention Calculations}: 
The calculation for a single head can be written as $\mathrm{Attention}(Q, K, V) = \mathrm{softmax}(\frac{QK^T}{\sqrt{d_k}})V$, where $d_k$ is the dimension of the queries and keys.

\textit{Aggregation}: The outputs from all heads are concatenated and projected back to the original dimension as
\begin{align*}
&\mathrm{MultiHeadAttention(Q, K, V) = \mathrm{Concat}(h_1, \dots, h_p)W^O},
\end{align*}
where $\medspace h_i = \mathrm{Attention}(QW^Q_i, KW^K_i, VW^V_i)$ and $p$ is the amount of heads.

The decoder's architecture is the same as the previously described encoder, but it includes an additional sub-layer that performs multi-head attention over the encoder's output. 
This extra layer enables the decoder to focus on the most relevant parts of the input sequence when generating the next token.

\subsection{Pre-Training}\label{sec:pre-training}

Large language models are pre-trained on enormous datasets that span trillions of tokens from diverse data sources such as books, academic articles, websites, code repositories, etc.
The main objective of pre-training is to guide the model to develop a rich, contextual understanding of language through self-supervised learning.

The training data for pre-training is unlabeled due to the sparse availability of labeled data.
There are multiple ways to train models on unlabeled data.
Masked Language Modeling, used to train BERT \cite{devlinBERTPretrainingDeep2019}, hides parts of the sequence and tasks the model by filling in a masked sequence based on the surrounding unmasked parts.
Casual language modeling and next-token prediction, popularized by the GPT models \cite{radfordImprovingLanguageUnderstanding}, involves training the model to generate the next token in a sequence. 
The loss of the training sample in these models is calculated as the difference between the generated tokens and the "true" tokens of the unlabeled training data. 
Qwen 2.5 Instruct, which is the base model we use, is trained on 18 trillion tokens of diverse training data \cite{qwenQwen25TechnicalReport2025}.

Since we utilize the pre-trained model in our work, we will not perform any pre-training steps for quantum circuit generation. However, the benefit we gain from a pre-trained model is that it is already capable of highly complex tasks and few-shot learning \cite{brownLanguageModelsAre2020}. We will further fine-tune this model, which will be discussed in the next section.

\subsection{Supervised Fine-Tuning}
\label{sec:sft}

While pre-training helps large language models gain a broad understanding of language, code, and reasoning through the enormous amounts of unlabeled data they are trained on, they do not inherently specialize in anything. The pre-trained models necessarily predict the next token in a sequence. Sequence generation is not usually favorable for human communication or more specific use cases like code generation.

This section introduces the concept of fine-tuning the already pre-trained model. Fine-tuning occurs during the subsequent training stage of a large language model, where the model is further trained on a smaller, often task-specific, labeled dataset in order to adapt the generalized knowledge embedded in the model for a more desired output.
The notable distinction between fine-tuning and pre-training is that pre-training is predominantly done on unlabeled data, while fine-tuning is done on curated, labeled data that is relevant to the purpose of fine-tuning.

There are several methodologies for fine-tuning large language models. \textit{Supervised Fine-Tuning} (SFT) is typically the first step post pre-training. 
SFT involves training the model on a dataset of curated input-output pairs, which often are formatted as instructions and their desired responses.
This input-output pairing makes the model learn to mimic the style, format, and behavior of these desired responses \cite{chowdheryPaLMScalingLanguage2022,touvronLLaMAOpenEfficient2023,ouyangTrainingLanguageModels2022}.
SFT can also be used to introduce new data, essentially making it learn new things in the context of its existing knowledge, to the model, which has been demonstrated well in the context of code generation \cite{roziereCodeLlamaOpen,deepseek-aiDeepSeekCoderV2BreakingBarrier2024,dupuisQiskitCodeAssistant2024}.

% In addition to SFT, Reinforcement Learning (RL) is widely used for fine-tuning models. 
% RL can be used during training to incentivize the model to follow specific behavior patterns \cite{ouyangTrainingLanguageModels2022} or to make the model converge on reasoning logic for complex logical problems \cite{deepseek-aiDeepSeekR1IncentivizingReasoning2025}.
% \textit{Reinforcement Learning from Human Feedback} (RLHF) can further align the model's behavior with trained human preferences.
% This process involves a separate \textit{reward model} that is trained on human preference data and then uses some RL algorithms like Proximal Policy Optimization \cite{schulmanProximalPolicyOptimization2017} to optimize the model's responses against the reward model.

\subsection{Supervised Fine-Tuning Pipeline for Circuit Generation}
\label{sec:sft-pipeline}

\begin{figure}
    \centering
    \includegraphics[width=0.99\columnwidth]{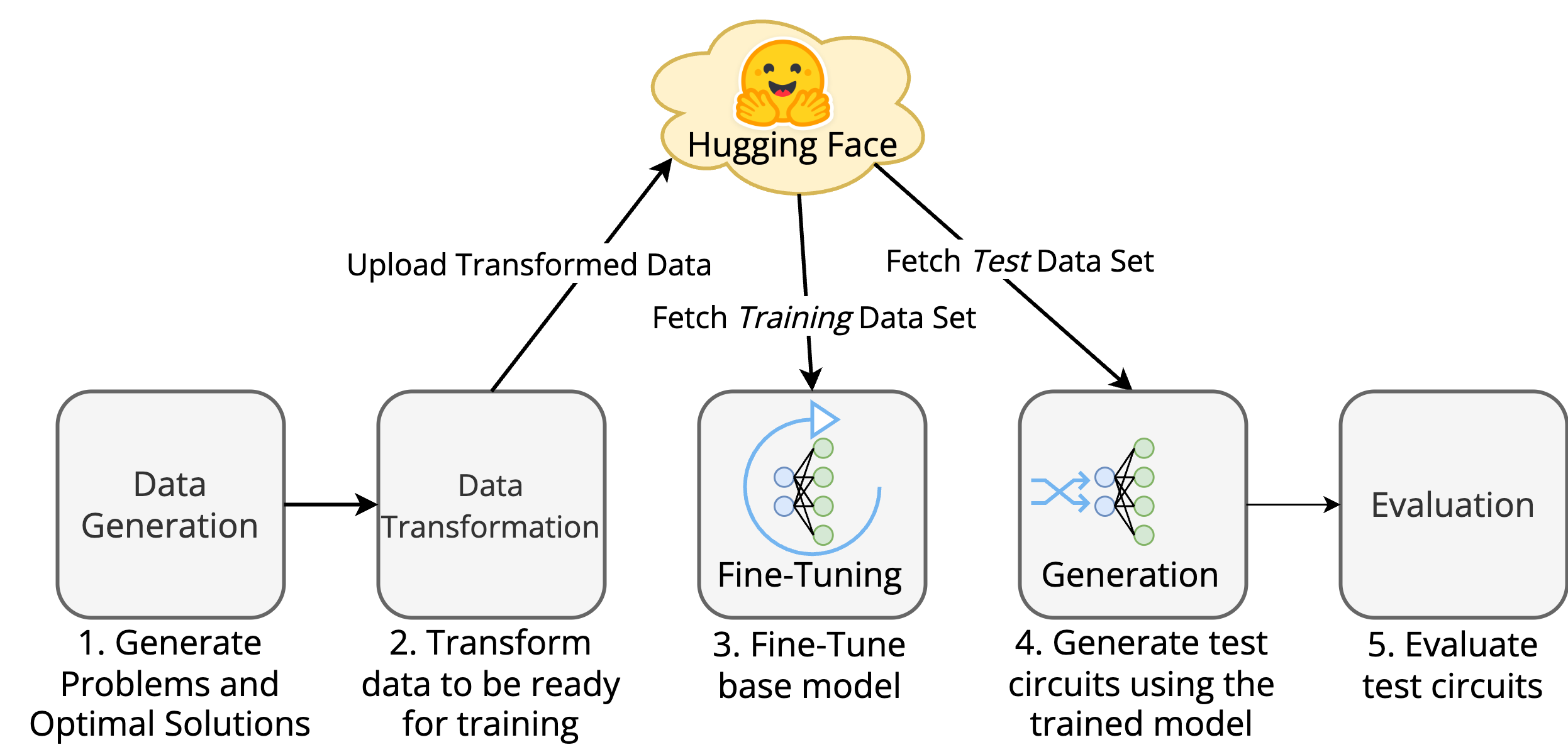}
    \caption{The end-to-end training pipeline of \sys}
    \label{fig:process_diagram}
\end{figure}

This section presents our fine-tuning pipeline, which makes large language models capable of generating contextually correct quantum circuits. 
For this purpose, we designed a dedicated SFT pipeline to fine-tune a general-purpose large language model for quantum circuit generation. 
This pipeline uses a pre-trained foundational model and further trains it on our specifically generated training data from the problems described in \autoref{sec:data}. 
Our pipeline uses the open-sourced pre-trained Qwen 2.5 Instruct model \cite{qwenQwen25TechnicalReport2025} as the base model.
This model was chosen due to its strong performance in code generation and instruction following tasks. The overall flow of the pipeline is based on a few core elements, visually illustrated in \autoref{fig:process_diagram}. 

\subsubsection{Data Generation}
\label{subsec:data_gen}

The foundation of any training process is a high-quality dataset. 
We use the dataset generated through the process presented in \autoref{sec:data} for our specific use case of fine-tuning a model for quantum circuit generation.
The optimization of the circuits is computationally heavy, and the data generation was thus conducted in a parallel manner over hundreds of processes.
The raw dataset pre-transformation can be found in \cite{linuzj_graph_data_quantum}.
One training sample contains all the relevant information necessary to build a training prompt from it and evaluate the generated circuit during evaluation. Our training dataset consists of 13,914 training samples. A train-test split of 96-4 is performed on this dataset. Although the 4\% test set is low by traditional ML standards, this split was chosen because it is reserved exclusively for computationally heavy post-training evaluation and is not consulted during the training phase itself.

\subsubsection{Data Transformation and Formatting}

\begin{figure}[t]
    \centering
    \includegraphics[width=0.92\columnwidth]{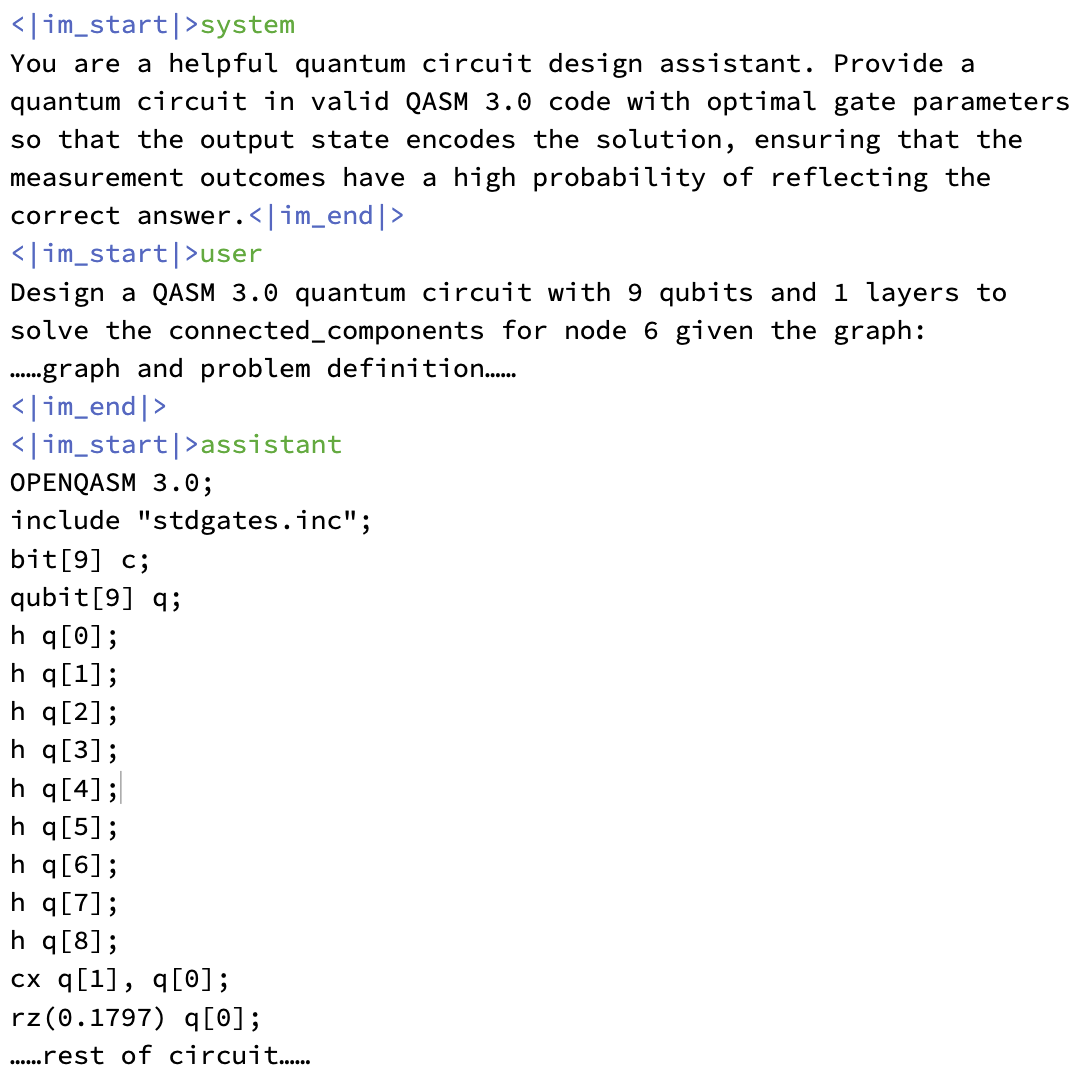}
    \caption{The snippet of one of the instruction prompts used during SFT. The blue strings are the model's special tokens that signal the start and end of sections. The green strings classify the sections.}
    \label{fig:prompt}
\end{figure}

The raw data points must be transformed into a format that is suitable for SFT.
This involves constructing a prompt from the relevant attributes that describes both the given problem and the optimized circuit, which serves as the label for the SFT training.
In our approach, we adopted a chat-based instruction template that formats the text into a string with special tokens, signaling to the model where the instruction begins and ends.
A sample snippet of this string is shown in \autoref{fig:prompt}.

Thus, one training sample is made up of one string, structured so that the training framework understands which part of said string is the \textit{prompt}, and which is the \textit{answer} (or label).
To further increase the variation in the data, we have five different prompt variants, each permutation having slightly different phrasing and ordering, which are randomly sampled during processing.
We have used Huggingface's \textit{Transformers} APIs \cite{wolfHuggingFacesTransformersStateoftheart2020} to build the training SFT pipeline.

\subsubsection{Supervised Fine-Tuning}

\begin{figure}[t]
    \centering
    \includegraphics[width=0.85\columnwidth]{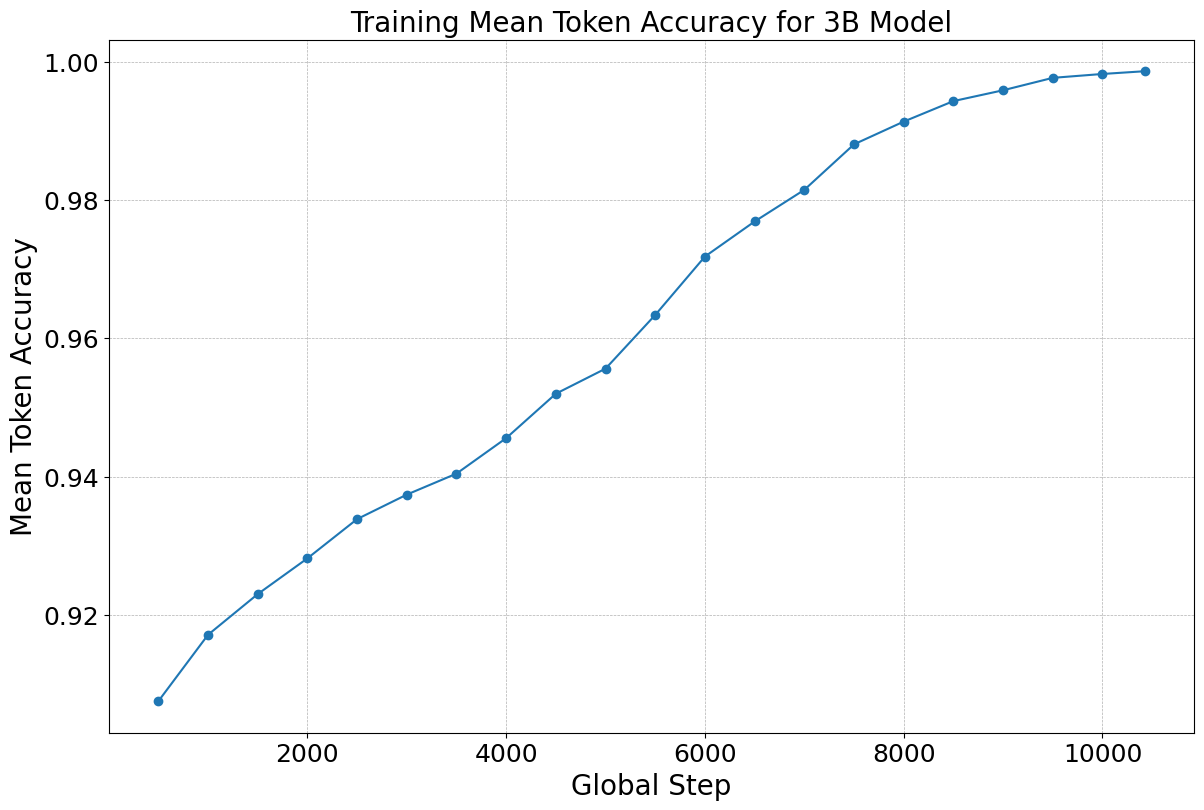}
    \caption{Training accuracy over global steps during training.}
    \label{fig:training_accuracy}
\end{figure}

The fine-tuning took 181 minutes using 5 Nvidia H200 GPUs.
We use simple fine-tuning hyperparameters, training for 15 epochs with a batch size of 1 per GPU and gradient accumulation steps of 4, resulting in a total of 10,425 optimization steps.
We train using BF16 floating-point precision and a learning rate of $2 \times 10^{-5}$.
The optmizer used is AdamW \cite{loshchilovDecoupledWeightDecay2019} using $\beta_1 = 0.99$, $\beta_2 = 0.999$ and weight decay of $1 \times 10^{-8}$.
A plot of the mean token accuracy during training is shown in \autoref{fig:training_accuracy}.

\subsubsection{Test Circuit Generation}
\label{subsec:generation}

Following the SFT training phase, the next step in the pipeline is to evaluate the performance of the model.
To evaluate the model's performance, a sample of test circuits needs to be generated based on data that the model has not yet seen.
The generated data, discussed in Section \ref{subsec:data_gen}, is divided into a \textit{training} split and a \textit{test} split.
The test split consists of 580 unique data points not present in the training data.
We sample 200 of these data points and use the fine-tuned model to generate quantum circuits used in evaluation.
\section{Evaluation}

We implement a comprehensive evaluation approach to systematically evaluate the performance of the fine-tuned model for quantum circuit generation.
We compare our model with other leading open-weight models: Llama 3.2 Instruct 3B \cite{grattafioriLlama3Herd2024}, DeepSeek R1 Distill 1.5B \cite{deepseek-aiDeepSeekR1IncentivizingReasoning2025}, and Gemma 3 4B \cite{teamGemma3Technical2025}.
Since we found that the general-purpose large language models struggle to produce any valid quantum circuits, we also include leading code generation models in our comparison: Qwen Coder 3B Instruct \cite{huiQwen25CoderTechnicalReport2024} and CodeGemma 7B \cite{teamCodeGemmaOpenCode2024}.
Furthermore, to fairly assess the capabilities of the previously mentioned models, we also evaluate all models using few-shot learning to give them QASM 3.0 syntax context \cite{brownLanguageModelsAre2020}.
Additionally, we compare the probability distributions of the generated circuits with those of the optimized circuits in the training data, as well as the expectation values of both circuits.

Our evaluation is structured around the following metrics:

\subsubsection{Syntatical Correctness}

A fundamental requirement of any model-generating code is the ability to produce syntactically correct code. Over larger outputs, this can be a challenging task for smaller models \cite{liu2023lostmiddlelanguagemodels}. In the context of quantum circuit generation, this means ensuring that the generated sequences are valid QASM 3.0 code. We assess the syntactical correctness of each generated circuit by parsing the generated sequences using a QASM 3.0 parser in Qiskit~\cite{javadi-abhariQuantumComputingQiskit2024}. A quantum circuit is deemed syntactically correct if it parses without errors, indicating that it follows all the grammatical rules of QASM 3.0. The results are displayed in \autoref{tab:syntatical_correctness}.

\begin{table}[t]
 \caption{Syntactical Correctness}
    \label{tab:syntatical_correctness}
    \centering
    \resizebox{\columnwidth}{!}{%
    \begin{tabular}{l r r}
    \toprule
    Model & \# Correct & Accuracy\\
      &  circuits &  rate \\
    \midrule
    \textbf{\sys} & \textbf{171} & \textbf{85.5} \% \\
    CodeGemma 7B + Few-Shot & 150 & 75 \% \\
    Llama 3.2 3B Instruct + Few-Shot & 122 & 61 \% \\
    DeepSeek R1 1.5B Distill 1.5B + Few-Shot & 57 & 28.5 \% \\
    Qwen 2.5 3B Instruct + Few-Shot & 49 & 24.5 \% \\
    Qwen 2.5 3B Coder Instruct + Few-Shot & 43 & 21.5 \% \\
    Qwen 2.5 3B Instruct & 0 & 0 \% \\
    Qwen 2.5 3B Coder Instruct & 0 & 0 \% \\
    DeepSeek R1 1.5B Distill 1.5B & 0 & 0 \% \\
    CodeGemma 7B & 0 & 0 \% \\
    \bottomrule
    \end{tabular}%
    }
\end{table}

\subsubsection{Expectation Value Analysis}\label{subsub:excp_val}

Given a syntactically correct quantum circuit, the natural question is whether the generated circuit represents the problem it was prompted to solve.
To answer this, we evaluate the expectation value of the generated quantum circuit with respect to the cost Hamiltonian of the problem.
In quantum optimization algorithms, the expectation value of a cost Hamiltonian effectively works as a value of the quality of the solution encoded in the quantum state \cite{farhiQuantumApproximateOptimization2014}. To assess the performance of our generated circuits, we use the following values based on expectation values:

\paragraph{Generated Expectation Value} Given a syntactically correct circuit, we simulate it using the Qiskit AerSimulator and then compute the expectation value of the problem-specific cost Hamiltonian. Let his value be $E_{\mathrm{gen}}$.

\paragraph{Solution Expectation Value} To establish a reference for comparison, we also calculate the expectation value of the optimized circuit from the test data set. Let this value be $E_{\mathrm{sol}}$.

\begin{figure}[t]
    \centering
    \includegraphics[width=0.99\columnwidth]{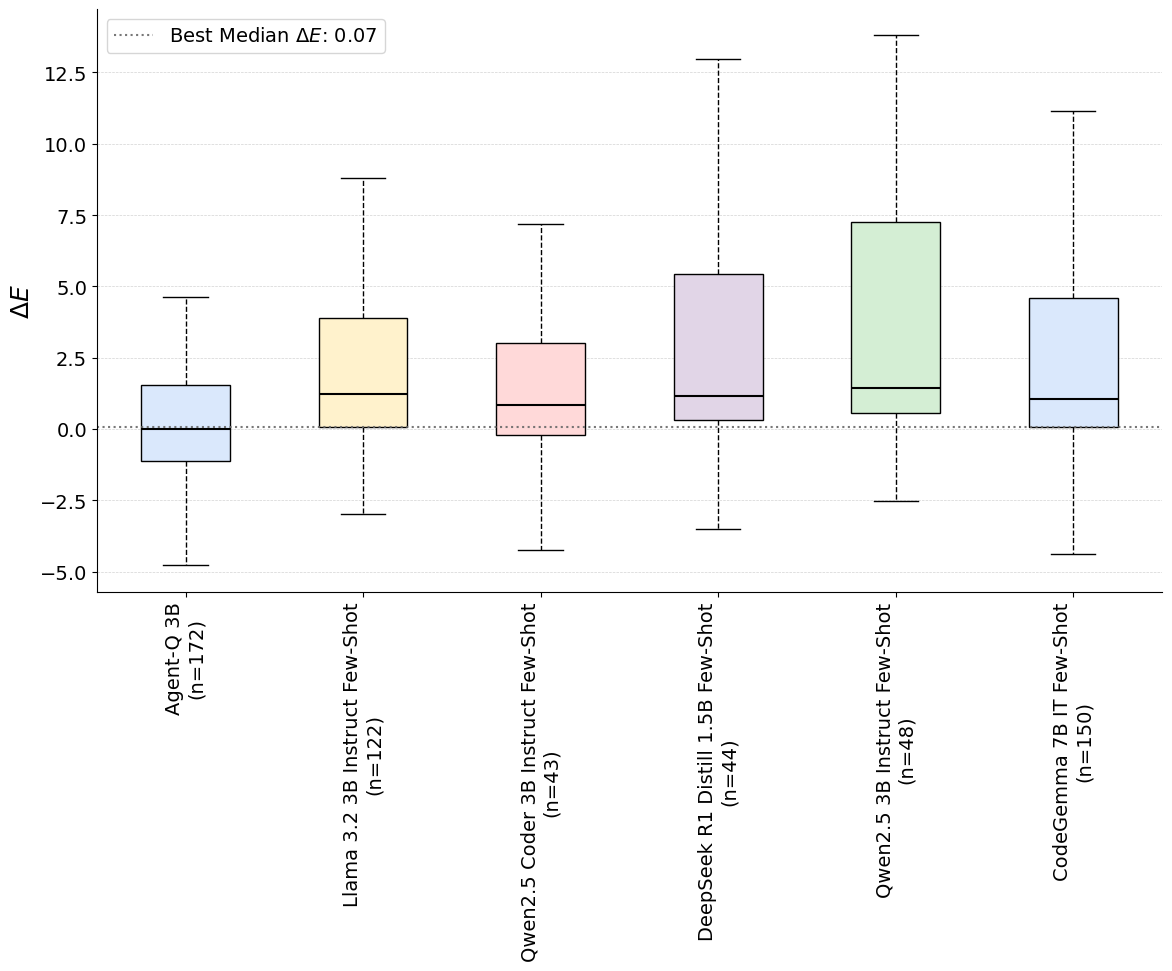}
    \caption{$\Delta E$ for successfully compiled circuits.}
    \label{fig:excpt_diff}
\end{figure}

\paragraph{Expectation Value Difference} To quantify the performance of the generated circuit, we calculate the absolute difference between the expectation values for the LLM-generated circuits and the expectation values from the solution circuits as: $\Delta E = |E_{\mathrm{gen}} - E_{\mathrm{sol}}|$. This metric includes information about the problem Hamiltonian, which is not included in the other metrics in this work. We interpret $\Delta E$ to mean that a smaller value indicates a higher-quality LLM-generated circuit. The values $E_{\mathrm{gen}}$ are computed for our model and the other pretrained models with few-shot learning.
We anticipate that well-generated circuit parameters should outperform the other models, given that the large language model has effectively generalized the problem structure. The Expectation Value Differences $\Delta E$ are displayed in \autoref{fig:excpt_diff}. The exact values are also shown in \autoref{tab:excpt_val_summary}.

\begin{table}[ht]
\caption{Summary statistics for the $\Delta E$ across models.}
    \label{tab:excpt_val_summary}
    \centering
    \resizebox{\columnwidth}{!}{%
        \begin{tabular}{l  r r r}
        \toprule
        Model Name & Mean $\Delta E$ & Median $\Delta E$& Std $\Delta E$\\
        \midrule
        \textbf{\sys}          & \textbf{0.53} & \textbf{0.07} & 8.01 \\
        Llama 3.2 3B Instruct Few-Shot          & 3.72 & 1.21 & 8.46 \\
        Qwen2.5 3B Coder Few-Shot           & 5.20 & 0.85 & 26.91 \\
        DeepSeek R1 Distill 1.5B Few-Shot         & 5.69 & 1.16 & 14.25 \\
        Qwen2.5 3B Instruct Few-Shot             & 6.14 & 1.44 & 14.66 \\
        CodeGemma 7B IT Few-Shot                   & 6.76 & 1.04 & 23.55 \\
        \bottomrule
        \end{tabular}
    }
\end{table}

\subsubsection{Relative Entropy of Probability Distributions}

Furthermore, we evaluate the performance of the generated circuits by calculating the similarity of the probability distributions of the generated circuits and the optimal solution circuit.
The output of a quantum circuit is fundamentally a probability distribution over measurement outcomes.
To argue that LLM-generated circuits have favorable characteristics, they should reproduce a probability distribution close to the target distribution of the optimal solution.

To quantify this, we calculate \textit{relative entropy}, also known as the \textit{Kullback-Leibler divergence} (KL)
\begin{equation*}
    D_{KL}(P \ || \ Q) = \sum_{x \in \mathcal{X}} P(x) \log \left( \frac{P(x)}{Q(x)}\right),
\end{equation*}
where $P$ and $Q$ are two probability distributions such that the support of $Q$ is a subset of the support of $P$. 
Relative entropy can be viewed as a measure of the distance between two probability distributions.
Given a generated circuit's probability $P_{\mathrm{gen}}$ and the optimal solution circuit's probability $P_{\mathrm{sol}}$, the relative entropy is $D_{KL}(P_{\mathrm{sol}} \ || \ P_{\mathrm{gen}})$.
Lower relative entropy indicates higher similarity between the two distributions.

In our evaluation, we calculate the relative entropy between the probability distributions obtained from simulating the generated circuit $P_{\mathrm{gen}}$, the optimal solution circuit $P_{\mathrm{sol}}$, and the baseline circuit with randomized parameters $P_{\mathrm{rand}}$ (same as in Section \ref{subsub:excp_val}). 
The results are displayed in \autoref{tab:relative_entropy_stats}.

\begin{table}[ht]
\caption{Average Relative Entropy ($D_{KL}$) values of the generated parameters compared to random parameters.}
    \label{tab:relative_entropy_stats}
    \centering
    \resizebox{0.7\columnwidth}{!}{%
        \begin{tabular}{ll}
        \toprule
        Metric & Value \\ \midrule
        Average $D_{KL}(P_{\mathrm{sol}} \ || \ P_{\mathrm{gen}})$  & 6.781 \\
        Average $D_{KL}(P_{\mathrm{sol}} \ || \ P_{\mathrm{rand}})$ & 9.623 \\
        \textbf{Improvement over random} & \textbf{29.5} \%
        \\
         \bottomrule
        \end{tabular}
    }
\end{table}
\section{Discussion} \label{sec:discussion}

The results show that the circuits' syntax can be learned efficiently, and the few-shot learning made the models comparable at the syntax level. OpenQASM syntax is relatively simple compared to natural language and programming languages. The recently published Google Gemma 7B performed the best with a few-shot tuning. Note that this model has 7B parameters, whereas we used the smallest Qwen model with 3B parameters.
These results are consistent with the observation that the competitor models lack the relevant quantum knowledge to perform well.

The fact that we focused on optimization problems provides us with the theoretical foundation to use the expectation values and relative entropy as evaluation metrics. Well-defined optimization problems have a limited set of optimal solutions that can be used in evaluation. Regarding the results from these two metrics, we note that our fine-tuned model outperformed all other state-of-the-art models. More importantly, we also found that the initial parameter values produced by the LLM produce lower expectation values than random guesses, and the corresponding distributions are closer to those measured from the optimized circuits. We view this finding as evidence that these circuits might be more efficient to optimize from the initial point given by the LLM model. This requires further experimental evaluation since optimization is dependent on multiple aspects. Nevertheless, the fine-tuned LLM model learned specific structures from the circuits and their parameters, and it is interesting to study further what these structures are.

We have identified multiple promising points to improve the model. First, the current implementation can be extended with reasoning based on the optimization process that the adaptive VQE method creates. Every step in adaptive VQE creates a circuit containing one more gate with a parameter to minimize the gradient. This leads to a sequence of circuits that could work as training data for reasoning models. 

Considering the circuit structure, using LLMs for quantum circuit compilation seems promising. This could be approached by the idea of ''translating circuits'' as natural language is translated. A logical circuit would correspond to the text in the source language, and the target text would be the compiled circuit, taking into account the fixed hardware topology.

Secondly, we are working on extending the model with reinforcement learning (RL).
Recently, impressive improvements in model performance for complex problems have been presented through the use of RL algorithms for fine-tuning, such as Group Relative Policy Optimization (GRPO) \cite{shaoDeepSeekMathPushingLimits2024}. 
We have also built an addition to the pipeline that implements GRPO for the quantum circuit generation process.
Due to its exploratory nature and the potential to define effective reward functions based on quantum simulation results (e.g., expectation value, circuit depth), GRPO has significant potential in both improving circuit parameters and developing novel circuit architectures.
GRPO-based circuits could potentially be tuned to compress circuit size, reducing both the size and depth of the circuits while maintaining expressivity.
This will be left for further research.

Furthermore, methods for model explainability are another direction of potential research.
Understanding why the model selects specific circuit sequences could reveal insights into its learned heuristics and logic, potentially leading to a deeper understanding of effective circuit construction.

Additionally, one consideration is the model's generalizability to various types of optimization problems.
Although the training data consists of a diverse set of 12 optimization problems, it has not yet been established how well the model performs on different optimization problems.
The current approach relies on natural language descriptions of the problem.
It could be possible to improve the generalization by encoding the problem more directly through its mathematical structure, for instance, by the cost Hamiltonian description. We also believe that the extensive publicly available training dataset of over 14,000 circuits will be useful for various tasks beyond fine-tuning LLM models.
\section{Conclusion and future work}

% \todo[color=green!30,inline]{Summary of key findings

% Concisely restate the main contributions and results

% Avoid repeating the abstract verbatim; instead, provide a high-level synthesis.
% }

% \todo[color=green!30,inline]{Significance \& implications

% Explain the broader impact of your work.

% Discuss practical applications, theoretical implications, or how they advance the field.
% }

% \todo[color=green!30,inline]{Limitations

% Acknowledge any constraints or assumptions in the study.
% }

% \todo[color=green!30,inline]{Future Work

% Suggest specific directions for extending your research

% Improving or refining your method.

% Applying the approach to different datasets, domains, or problems.

% Addressing any limitations.

% Exploring alternative models, techniques, or theoretical frameworks.
% }

%This article focused on optimization problems. Naturally, there are countless other use cases for parameterized quantum circuits, such as quantum machine learning.

%\textcolor{red}{Many of the future research possibilities were already discussed in the discussion section. The main next step is to include reinforcement learning as a part of the fine-tuning pipeline.}

\subsection{Conclusion}
In this study, we presented a fine-tuning approach based on Supervised Fine-Tuning (SFT) for large language models specifically tailored for quantum optimization tasks, demonstrating their potential to effectively generate parametrized quantum circuits and suitable initial parameters. Our fine-tuned model significantly outperformed baseline state-of-the-art language models, achieving high syntactical correctness and generating initial parameters closer to optimal values than random initialization. Furthermore, the generated circuits exhibited probability distributions that were considerably closer to optimal solutions as measured by relative entropy.

The significance of this work lies in its practical applications in quantum computing. Our model can assist quantum algorithm developers by providing strong starting points for optimization routines, thus accelerating quantum algorithm development and potentially enhancing the efficiency of quantum computation processes. Furthermore, the produced circuits can serve as benchmarks for quantum compilers and quantum hardware evaluations, marking a step toward more sophisticated hybrid quantum-classical programming frameworks.

\subsection{Limitation}
Our current model's generalization capability to entirely unseen quantum optimization problems has not been fully assessed. The diversity and complexity of the optimization problems used for training, although extensive, may still restrict the model's performance on vastly different quantum computational tasks. Additionally, the complexity of quantum computations inherently limits current simulation-based evaluation methods to relatively small quantum systems; we still need to explore more efficient methods for large-scale quantum circuit verifications.
Furthermore, because the training and evaluation sets draw on the same problem classes, the strong metrics reported may partially stem from overfitting to recurring structures rather than from task-agnostic learning.

\subsection{Future Work}
Future research directions, as discussed in detail in the preceding section, include integrating reinforcement learning approaches such as Group Relative Policy Optimization (GRPO) into the fine-tuning process. These methods could enhance parameter selection and lead to novel circuit structures optimized for quantum hardware constraints. Future research could focus on using explainability tools to better understand model decisions, integrating mathematical concepts like Hamiltonian encodings into training prompts, and expanding into other areas of quantum computing.

\section{Acknowledgement}
This work is funded by Business Finland (grant number 169/31/2024), Research Council of Finland (grant number 362729) and the Quantum Doctoral Programme. We also acknowledge the computational resources provided by the Aalto Science-IT project.

\balance
\bibliographystyle{IEEEtran}
\bibliography{ref.bib}

\end{document}